# Surface mobility of a glass-forming polymer in an ionic liquid

Xinyu Zhang,[1] Christian Pedersen,[3,4] Haoqi Zhu,[1] Siming Wang,[1] Yuchen Fu,[1] Liang Dai,[1] Andreas Carlson,[3,5] Thomas Salez,[2,*] and Yu Chai [1,*]

[1] *Department of Physics, City University of Hong Kong, 83 Tat Chee Avenue, Kowloon, Hong Kong SAR, China.*
[2] *Univ. Bordeaux, CNRS, LOMA, UMR 5798, Talence F-33400, France.*
[3] *Mechanics Division, Department of Mathematics, University of Oslo, 0316 Oslo, Norway.*
[4] *Expert Analytics AS, N-0179 Oslo, Norway.*
[5] *Department of Medical Biochemistry and Biophysics, Umeå University, Umeå, Sweden.*
*Corresponding authors: yuchai@cityu.edu.hk; thomas.salez@cnrs.fr

The free surface of glassy polymers exhibits enhanced segmental dynamics compared to the bulk, forming a liquid-like layer that lowers the glass transition temperature ($T_g$) in nanometer-sized polymer samples. Recent studies have shown that immersing polymers in ionic liquids can suppress this enhanced surface dynamics. To investigate how ionic liquids influence polymer dynamics near the ionic-liquid-polymer interface, we measure the surface leveling of nanometer-sized stepped polystyrene films immersed in ionic liquids, and compared the results to the case of films in vacuum. Our results reveal that ionic liquids significantly slow the leveling process both above and below $T_g$. However, our results indicate that the liquid-like surface layer below $T_g$ does exist in ionic liquids. Numerical solutions of the thin-film equation, incorporating appropriate boundary conditions, show that the surface mobility of PS films in ionic liquids can match that of PS films in vacuum. Thus, while ionic liquids alter the polymer flow process, they do not eliminate the dynamical heterogeneity inherent to glassy polymers.

Thin polymer films often exhibit a reduced glass-transition temperature ($T_g$) when their thickness is below ~ 40 nm [1], primarily due to a free-surface effect. Indeed, polymer chains near the free surface (polymer-air or polymer-vacuum interface) experience fewer free-volume and enthalpic constraints, thus enhancing molecular dynamics and lowering the local $T_g$ [2]. This results in the formation of a liquid-like layer even when the bulk remains glassy. As the film thickness decreases, the increased surface-to-volume ratio amplifies such an effect [3,4]. Studies have shown that coating the polymer film with a solid layer can eliminate $T_g$ reductions by suppressing the free-surface effect. Strong interactions between polymers and substrates can even increase $T_g$. For example, in polymethyl methacrylate (PMMA) on a silicon (Si) wafer, strong interactions raise $T_g$ as the film thins [5]. This demonstrates how $T_g$ in thin polymer films is governed by competing interfacial effects.

More recently, attention has shifted from the study of polymers confined to free surfaces [6,7], or hard substrates [1,5], to investigating their behavior in more complex environments, such as liquid atmospheres. Among various liquids, ionic liquids have shown the ability to suppress the free surface effect and eliminate the $T_g$ reduction in thin polymer films. For example, as illustrated in Fig. 1(a), polystyrene (PS) —whether in the form of thin films or nanoparticles— maintains bulk-like $T_g$ values when in contact with ionic liquids [8,9], specifically 1-butyl-3-methylimidazolium trifluoro-methanesulfonate ([BMIM][CF$_3$SO$_3$]). This behavior is different from polymers with free surfaces [1,10,11], or those in contact with water [12,13], where $T_g$ still decreases as the typical sample size is reduced.

The ability of ionic liquids to strongly influence the $T_g$ of polymers at the nanometer scale raises an important question: how do ionic liquids alter the molecular or segmental dynamics of polymers near the interface? To address this, we conduct stepped-film leveling experiments in ionic atmospheres, to directly probe the dynamics of polymers near the polymer-ionic-liquid interface. Our results reveal that ionic liquids slow down the curvature-induced capillary leveling of stepped PS thin films. However, analysis of the film profiles extracted using atomic force microscopy (AFM) confirms that a liquid-like surface layer [15,16] still exists below $T_g$ when the polymer film is in contact with ionic liquids. By modifying the boundary conditions in thin-film models based on lubrication theory, we find that polymers exhibit similar surface mobilities regardless of whether the glassy films are in contact with ionic liquids or vacuum. This suggests that while ionic liquids slow down the flow of polymers near the ionic-liquid-polymer interface, through a frictional hydrodynamic boundary condition, they do not eliminate the dynamical heterogeneity inherent to glassy polymers due to the free surface effect.

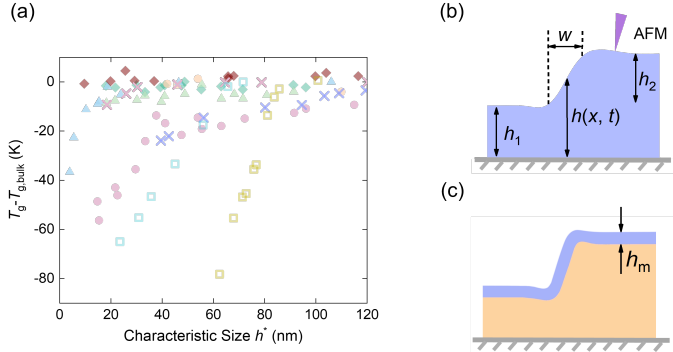

FIG. 1. (a) The shift in the glass transition temperature with respect to the bulk one $T_g$ - $T_{g,bulk}$, as a function of the characteristic system size $h^*$, which can be either the thickness of polymer films or the diameter of nanoparticles, in various atmospheres. (◆) Nanoparticles dispersed in [BMIM][CF$_3$SO$_3$] and (▲) glycerol [8]; (●, ○) Nanoparticles dispersed in water [12,13]; (▲) Thin films floating on glycerol [14]; (◆) Thin films floating on [BMIM][CF$_3$SO$_3$] [9]; (×, ×) Supported thin films in air [1,10]; (□, □) Freestanding films in air [11]. (b-c) Schematic diagrams of the flow mechanisms in silicon-supported stepped films. The height profile $h(x,t)$ for: (b) whole-film flow, and (c) near-surface flow. The mobile region is indicated in blue, while the immobile region is indicated in orange.

## RESULTS AND DISCUSSION

To conduct the stepped-film leveling experiments, we first fabricate the stepped polymer films [17]. The films consist of a bottom PS layer with thickness $h_1$ (80 ~ 320 nm) and a top PS layer with the same thickness $h_2$ (see Fig. 1(b)). The bottom layer is spin-coated onto a clean silicon substrate from a dilute PS solution in toluene (Polymer Source, $M_w$ = 2.481 kg/mol, polydispersity index = 1.08). Similarly, the top layer is spin-coated onto freshly cleaved mica. Both PS films (on silicon and mica substrates) are pre-annealed in a vacuum at 80 °C for at least 24 hours to remove residual stresses and solvent. After annealing, the top PS film (on mica) is floated onto ultra-pure water and carefully transferred onto the bottom PS-coated silicon substrate. During this process, the top PS layer fractures into small pieces with straight vertical edges upon perturbation at the water surface, creating well-defined steps upon transfer. The thickness of the films and the glass-transition temperature of PS ($T_g$ ≈ 64.4 ± 2 °C) are measured using ellipsometry (J.A. Woollam, RC2). The films studied in this paper are thick enough so the measured $T_g$ remains the bulk value. After preparation, all stepped PS samples are carefully stored before the leveling experiments are conducted.

For leveling experiments, the stepped PS films are annealed in two distinct environments: i) in a vacuum chamber; or ii) immersed in a vial filled with 1-butyl-3-methylimidazolium trifluoromethanesulfonate ([BMIM][CF$_3$SO$_3$]). The samples are annealed for predetermined time periods in their respective environments and measured using AFM (Cypher ES, Oxford Instruments) under ambient conditions. This process is repeated to observe the temporal evolution of the stepped PS films in different environments.

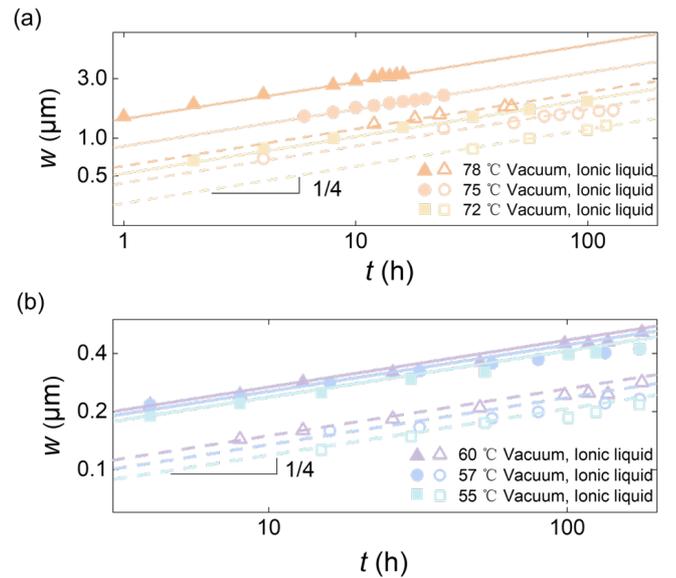

FIG. 2. Width $w$ of stepped PS films as a function of time $t$, for various temperatures and atmospheres. (a) Temporal evolution of the width $w$ (shown in Fig. 1), obtained by fitting the profile to a $tanh[x/(w/2)]$ function, for $h_1 = h_2 = 180 \pm 5$ nm, at temperatures above the bulk $T_g$. (b) Temporal evolution of the width, for films with $h_1 = h_2 = 78 \pm 3$ nm, at temperatures below the bulk $T_g$. Data for ionic-liquid and vacuum atmospheres are presented by hollow symbols fitted with dashed lines and solid symbols fitted with solid lines, respectively. All dashed and solid lines correspond to power laws with an exponent of 1/4.

As annealing time increases, regardless of whether the system is above or below $T_g$, in vacuum or immersed in ionic liquids, the stepped PS films all exhibit flow, as evidenced by the systematic broadening of the step. To quantify the temporal evolution of the stepped films, we fit the experimental height profiles $h(x,t)$ of the stepped films at given times $t$ to the ad-hoc $\tan[x/(w/2)]$ functional form, where $w(t)$ is the time-dependent width. Fig. 2 shows the width of the stepped PS film as a function of time at different temperatures, in two environments. The results reveal that the width always scales with time following a power law with an exponent of 1/4, but the evolution is noticeably slower for films immersed in ionic liquids compared to those in vacuum. However, this slowing down does not necessarily suggest that ionic liquids eliminate the liquid-like surface layer below $T_g$, as they also slow down the leveling above $T_g$, thus the bulk flow. Therefore, to better understand the effects of ionic liquids on the leveling of stepped polymer films, it is of great importance to invoke appropriate physical models.

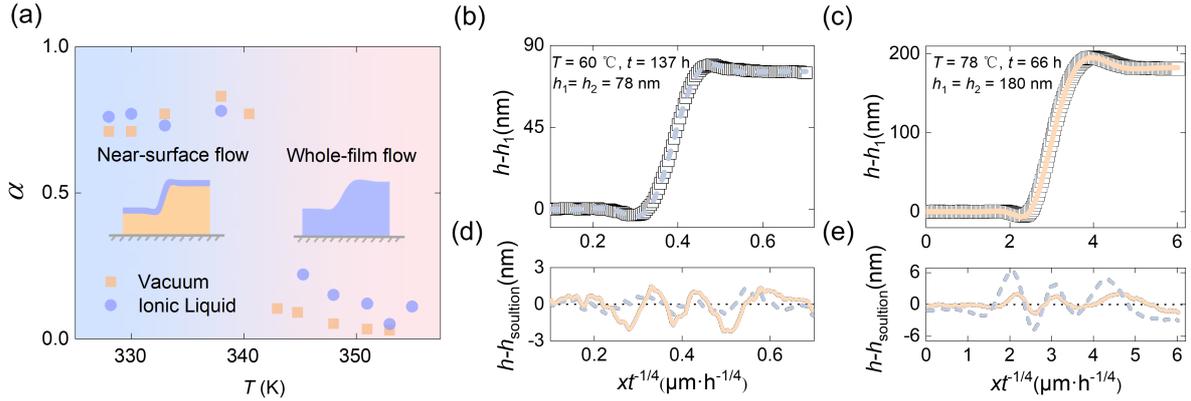

FIG. 3. (a) Correlation function $\alpha$ (see definition in text) as a function of temperature $T$, in both vacuum and ionic-liquid atmospheres. The correlation function $\alpha(T)$ is represented by orange squares (vacuum) and purple circles (ionic liquid) for samples with equal $h_1$ and $h_2$ of 50 nm, 78 nm, 180 nm, and 290 nm. The insets show schematic diagrams of surface flow (left) and whole-film flow (right), as in Fig. 2. (b) and (c) Experimental profiles (white squares) of glassy (b) and melt (c) PS films in an ionic liquid, fitted to the GTFE model (blue dashed line, see definition in text) and the TFE model (orange solid line, see definition in text). (d) and (e) Difference between best-fit theoretical and experimental profiles as a function of the self-similar variable $xt^{-1/4}$, for the GTFE solution (blue dashed line) and the TFE solution (orange solid line), below $T_g$ (d) and above $T_g$ (e).

Previous works established that the leveling process of stepped films in air or vacuum can be described by two partial differential equations: the thin film equation (TFE) [18,19] for temperatures above $T_g$, and the glassy thin film equation (GTFE) [15] for temperatures below $T_g$. In the TFE model, the entire film is assumed to be a Newtonian liquid with a homogeneous viscosity ($\eta_b$), and undergoes bulk flow. The leveling process is mathematically described by the TFE:

$$\frac{\partial h}{\partial t} + \frac{\gamma}{3\eta_b}\frac{\partial}{\partial x}\left(h^3 \frac{\partial^3 h}{\partial x^3}\right) = 0, \qquad (1)$$

where $\gamma$ is the PS-vacuum (or PS-air) surface tension. In contrast, the GTFE model assumes that only a surface layer of thickness $h_m$ is liquid and capable of flow, with viscosity $\eta_m$ [20], while the remainder of the film is solid and immobile. This surface-flow-driven leveling process is mathematically described by the GTFE:

$$\frac{\partial h}{\partial t} + \frac{\gamma h_m^3}{3\eta_m}\frac{\partial^4 h}{\partial x^4} = 0, \qquad (2)$$

whose solution can be obtained analytically [21].

Although both the TFE and GTFE are fourth-order diffusive-like partial differential equations, they predict distinct leveling profiles. Due to the asymmetric nature of the stepped films, the non-linear TFE predicts a bigger bump in the profile on the higher step side compared to the dip on the lower step side. In contrast, for the linear GTFE, where the liquid mobile layer $h_m$ has a constant thickness, the bump and dip have the same magnitude. These results have been previously confirmed and recovered in the case of stepped films annealed in vacuum in the present study.

Interestingly, even though ionic liquids slow down the leveling process of the stepped films, the experimental profiles obtained by AFM can still be perfectly described by the numerical solution of the TFE [16] for the above-$T_g$ case and by the analytical solution of the GTFE [21] for the below-$T_g$ case, as shown in Figs. 3(b)-(e). The results undoubtedly imply that ionic liquids do not remove the liquid-like surface layer observed in glassy PS.

To quantify how closely our experimental profiles fit either the numerical profile calculated from the TFE or the analytical profile calculated from the GTFE, we introduce a correlation function $\alpha$, defined as:

$$\alpha = \frac{\int dx \, (h_{\exp} - h_{\text{TFE}})^2}{\int dx \, (h_{\exp} - h_{\text{GTFE}})^2 + \int dx \, (h_{\exp} - h_{\text{TFE}})^2}, \quad (3)$$

where $h_{\exp}(x)$ is the experimental profile, and $h_{\text{TFE}}(x)$ and $h_{\text{GTFE}}(x)$ are the theoretical profiles computed from the TFE and GTFE respectively. If $\alpha$ is equal to 1, the experimental profiles are best described by the GTFE, while if $\alpha$ is 0, the experimental profiles are best described by the TFE. Fig. 3(a) shows the transition of $\alpha$ from 1 to 0 with increasing temperature for stepped films annealed both in vacuum and ionic liquids, with the transition temperature close to the bulk $T_g$ value.

We now turn to understanding why ionic liquids slow the leveling process of stepped PS films, both above and below $T_g$, without affecting the transition between surface-flow-driven leveling below $T_g$ and bulk-flow-driven leveling above $T_g$. Above $T_g$, the leveling process is dominated by the bulk dynamic shear viscosity ($\eta_b$) of the viscous molten film. The observation that ionic liquids slow the leveling process even above $T_g$ raises the possibility of long-range Van der Waals interactions between the ionic liquid and the Si substrate. To test this hypothesis, we include an additional disjoining pressure in the excess pressure field $p$ at the free interface of the film, as:

$$p_{\text{vdW}}(x,t) = -\gamma \frac{\partial^2 h}{\partial x^2} + \frac{A}{6\pi h^3}\left[1 - \left(\frac{\delta}{h}\right)^6\right], \quad (4)$$

with $A$ the Hamaker constant and $\delta$ the sixth power of the equilibrium thickness at which the attractive and repulsive pressure terms cancel each other [22]. By incorporating this modified excess pressure field $p$ into the TFE and GTFE, we derived modified versions of both models that account for the influence of van der Waals interactions. We then numerically solved non-dimensional versions of the modified equations for $A$ in [0, 1×10$^{-17}$ N·m] and $\delta$ in [0, 7×10$^{-8}$ m] as detailed in the SI.

The results indicate that if such long-range interactions existed, their magnitude would have to be sufficiently large to account for the observed differences in leveling between vacuum and ionic-liquid environments. However, such strong van der Waals interactions would likely lead to significant deviations from the 1/4 scaling exponent. Therefore, we rule out this hypothesis.

The derivation of the TFE and GTFE is based on two flow boundary conditions: the liquid-atmosphere interface is assumed to be without shear, while the polymer-substrate interface or the polymer (surface liquid layer)-polymer (bulk glassy layer) interface is assumed to be without slip. However, when the environment changes from air or vacuum to an ionic liquid, the boundary condition may also change.

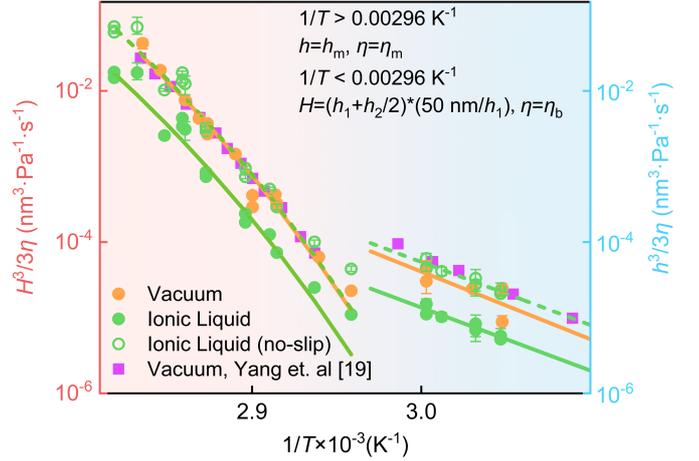

FIG. 4. Mobility $H^3/(3\eta)$ of thin polymer films as a function of inverse temperature $1/T$, in both vacuum (orange circles) and ionic liquids, with either a no-shear boundary condition (solid green circles) as in vacuum or a no-slip boundary condition (hollow green circles) at the ionic-liquid-polymer interface. The results of Yang et al. [19] in vacuum are also plotted as purple squares for comparison. The bulk mobility (corresponding to the left y-axis) $H^3/(3\eta)$ for $T > T_g$ is determined by fitting the numerical TFE solution to the experimental profiles. The results are compared to the Vogel-Fulcher-Tammann law (solid and dashed curves on the left side). The surface mobility (corresponding to the right y-axis) $h^3/(3\eta)$ for $T < T_g$ is obtained by fitting the analytical GTFE solution to the experimental profiles. The results are compared to the Arrhenius law (solid and dashed lines on the right side). $H$, $h$, and $\eta$ are defined in the legend for the two temperature regions.

Considering the short-range (contact) interactions between ionic liquids and polymers, the no-shear boundary condition at the polymer-atmosphere interface is now replaced with a no-slip boundary condition in the ionic-liquid case. This modification leads to two new versions of the TFE and GTFE, termed TFE-no-slip and GTFE-no-slip, respectively:

$$\frac{\partial h}{\partial t} + \frac{\gamma}{12 \eta_b} \frac{\partial}{\partial x}\left(h^3 \frac{\partial^3 h}{\partial x^3}\right) = 0, \quad (5)$$

$$(6)$$

$$\frac{\partial h}{\partial t} + \frac{\gamma h_m^3}{12\eta_m}\frac{\partial^4 h}{\partial x^4} = 0 \,.$$

The only difference between these no-slip versions and the original equations is the numerical prefactor in the mobility, which changes from 1/3 to 1/12. Using this new numerical factor and the interface tension $\gamma_{\text{PS-Ionic Liquid}}$ = 8.74 mN/m, we found that the mobilities of the PS films in vacuum and ionic-liquid environments are the same in both above and below $T_g$ cases. Fig. 4 assembles the mobility data of PS films in various environments, with stepped films of different thicknesses ($h_1$ and $h_2$, scaled to 50 nm for direct comparison). For reference, the mobility data for films in air, obtained from surface-roughening experiments by Yang et al. [23] using a similar molecular weight of PS, are also shown. It is evident that all data sets overlap. By fitting the below-$T_g$ mobility with the Arrhenius law, we find activation energies on the order of $E_a \approx$ 171.25 kJ/mol for the vacuum case and $E_a \approx$ 159.69 kJ/mol in ionic liquids. These values are in good agreement, within 20%, with the activation energy of 185 kJ/mol reported by Yang et al. [23]. It should be noted that assuming a complete no-slip boundary condition at the ionic-liquid interface provides an upper bound for the estimated mobility values.

## CONCLUSIONS

In summary, we measured the mobility of thin polystyrene films in two different environments: vacuum and ionic liquids, using stepped-film leveling experiments. Our results show that ionic liquids significantly slow down the flow-driven leveling process, both above and below $T_g$. However, the experimental profiles suggest that ionic liquids do not suppress the liquid-like surface layer in glassy PS, below $T_g$. By incorporating the appropriate flow boundary conditions into the governing thin-film equations, we found that the mobility values are similar in both vacuum and ionic-liquid environments, both above and below $T_g$.

## ACKNOWLEDGMENTS

The authors thank James Forrest for insightful discussions. They acknowledge financial support from the Research Grants Council of Hong Kong (No. 21304421, Y.C.); the National Natural Science Foundation of China (No. 22003053, Y.C.); the Natural Science Foundation of Guangdong Province, China (No. 2023A1515011457, Y.C.); the Natural Science Foundation of Sichuan Province, China (No. 2023NSFSC0312, Y.C.); and the CityU Strategic Interdisciplinary Research Grant (No. 2020SIRG035, Y.C.). They also acknowledge support from European Union through the European Research Council under EMetBrown (ERC-CoG-101039103) grant. Views and opinions expressed are however those of the authors only and do not necessarily reflect those of the European Union or the European Research Council. Neither the European Union nor the granting authority can be held responsible for them. The authors also acknowledge financial support from the Agence Nationale de la Recherche under Softer (ANR21-CE06-0029) and Fricolas (ANR-21-CE06-0039) grants, as well as from the Interdisciplinary and Exploratory Research Program under MISTIC grant at the University of Bordeaux, France. Finally, they thank the RRI Frontiers of Life, which received financial support from the French government in the framework of the University of Bordeaux's France 2030 program; as well as the Soft Matter Collaborative Research Unit, Frontier Research Center for Advanced Material and Life Science, Faculty of Advanced Life Science, Hokkaido University, Sapporo, Japan; and the CNRS International Research Network between France and India on "Hydrodynamics at small scales: from soft matter to bioengineering". The authors also thank OpenAI for editing and proofreading assistance during the preparation of the manuscript.